\documentclass[12pt]{article}
\usepackage{epsfig}

\usepackage{amssymb}
\usepackage{amsmath}
\usepackage{amsfonts}
\usepackage{amsthm}

\theoremstyle{theorem}
\newtheorem{lem}{Lemma}[section]

\newtheorem{prop}[lem]{Proposition}
\newtheorem{thm}[lem]{Theorem}
\newtheorem{coro}[lem]{Corollary}

\theoremstyle{definition}

\newtheorem{pozn}[lem]{Remark}

\def\bp{\begin{proof}}
\def\ep{\end{proof}}

\def\be{\begin{equation}}
\def\ee{\end{equation}}
\def\ba{\begin{array}{c}}
\def\ea{\end{array}}

\def\ben{$$}
\def\een{$$}

\newcommand{\bea}{\begin{eqnarray}}
\newcommand{\eea}{\end{eqnarray}}

\newcommand{\bbr}{\br\!\br}

\newcommand{\kt}{\rangle}
\newcommand{\br}{\langle}
\begin{document}

\titlepage

\vspace{.35cm}

 \begin{center}{\Large \bf

Complete set of inner products for a discrete ${\cal PT}-$symmetric
square-well Hamiltonian

  }\end{center}

\vspace{10mm}

 \begin{center}

 {\bf Miloslav Znojil}

 \vspace{3mm}
Nuclear Physics Institute ASCR,

250 68 \v{R}e\v{z}, Czech Republic

{e-mail: znojil@ujf.cas.cz}

\vspace{3mm}


\end{center}

\vspace{5mm}


\section*{Abstract}

A discrete $N-$point Runge-Kutta version $H^{(N)}({\lambda})$ of one
of the simplest non-Hermitian square-well Hamiltonians with real
spectrum is studied. A complete set of its possible hermitizations
(i.e., of the eligible metrics $\Theta^{(N)}({\lambda})$ defining
its non-equivalent physical Hilbert spaces of states) is
constructed, in closed form, for any coupling ${\lambda}\in (-1,1)$
and any matrix dimension $N$.


\newpage
 \section{Introduction \label{zacatek} }

\subsection{Bound states in Runge-Kutta approximation}

The concept of the solvability of a dynamical model in physics is
rather vague. Its definition is usually adapted to the range of
expected applications. By some authors even the single-particle
motion along a finite one-dimensional interval would be called
solvable only if the underlying ordinary differential
Schr\"{o}dinger equation for bound states
 \be
 -\frac{d^2}{dx^2}\,\psi(x)+
  V(x)\,\psi(x) =E\,\psi(x)\,,
 \ \ \ \ \ \ \
 \psi(\pm L)=0
 \label{SEloc}
 \ee
proved reducible to the Gauss' or confluent hypergeometric equation.
Remarkably enough, even this extremely narrow specification of
solvability finds very plausible physical reasons in the related
shape invariance of potentials $V(x)$ and/or in their close
relationship to supersymmetry \cite{Cooper}.

In what follows we shall rather accept different terminology in
which one treats virtually {\em any} sufficiently smooth and real
potential $V(x)$ in Eq.~(\ref{SEloc}) as exactly solvable, i.e.,
solvable, in principle, with arbitrary precision, in a purely
numerical setting at least. Typically, the Runge-Kutta (RK,
\cite{thatwork}) discrete approximation of the above equation, viz.,
 \be
 -\frac{\psi(x_{k-1})-2\,\psi(x_k)+\psi(x_{k+1})}{h^2}+V(x_k)\,
 \psi(x_k)
 =E\,\psi(x_k)\,,
 \label{SEdis}
 \ee
 \ben
 \ \ \ \ \ \ \ \ \
 \ \ \ \ x_k=k\,h\,, \ k = 0, \pm 1, \ldots,
 \pm K\,,
 \ \ \ \ x_{\pm K}=\pm L\,,\ \ \ \ \ \
 \psi(x_{\pm (K+1)})=0\,
 \een
may be used to reduce the original bound-state problem to a routine
computer-assisted diagonalization of the finite-dimensional RK
matrix Hamiltonian
 \be
 H=
 \left (
 \begin{array}{ccccc}
 \ddots &\ \ \ \ \ \ \ \ \ \ \ddots &&& \\
 \ddots &2+h^2V(x_{-1})&-1&&\\
 &-1&2+h^2V(x_{0})&-1\\
 &&-1&2+h^2V(x_1)&\ddots\\
  &&& \ddots\ \ \ \ \ \ \ \ \ \ & \ddots
 \ea
 \right )\,.
 \label{dvanact}
 \ee
The choice of the lattice distance $h\ll 1$ is only dictated by the
required precision of reproduction of the original energies and/or
wave functions.

\subsection{${\cal PT}-$symmetric Runge-Kutta models}

The standard formalism of quantum theory  admits Schr\"{o}dinger
Eqs. (\ref{SEloc}) and/or (\ref{SEdis}) which generate the {\em
real} bound-state spectra from certain {\em complex} potentials
$V(x)$. Conventionally, these potentials are called ${\cal
PT}-$symmetric (cf. reviews \cite{Carl,Carl2} or Appendix A for more
details). For a sensible extension of the concept of solvability to
this new context it is most important that the ${\cal PT}-$symmetric
models are characterized by the Hamiltonian-dependence of their
physical Hilbert spaces ${\cal H}^{(S)}$. Although the superscript
$^{(S)}$ stands here for ``standard", the inner product is defined
in a nonstandard manner in this space,  by a formula containing an
{\em ad hoc} metric operator $\Theta=\Theta(H)\neq I$ (cf.
Eq.~(\ref{innerie}) in Appendix A below).

In spite of the existence of several powerful techniques of
reconstruction of $\Theta$ for differential Eq.~(\ref{SEloc})
\cite{Carl,cubic} one can rarely find a closed-form result (for
illustration check a few samples in
Refs.~\cite{swansonfring,david}). Many successful constructions rely
upon various assumptions requiring, e.g., the existence of a charge
of the system \cite{BBJ,plb} or of some other additional and/or
complementary observable(s) ${\cal O}$ \cite{Geyer}. Moreover,
people are mostly able to obtain $\Theta$ just in an approximate
form, say, of perturbation series \cite{Bataldelta}.

Some of these difficulties have been addressed in our papers
\cite{ojonesovi}. We restricted our attention to the discrete
Schr\"{o}dinger Eq.~(\ref{SEdis}) considered at a pre-determined
precision, i.e., at a fixed spacing constant $h>0$. This enabled us
to broaden the scope of the theory and to consider certain nonlocal
generalizations of interaction terms. More explicitly, we
complemented the diagonal elements $V(x_k)$ in the RK matrix
Hamiltonian (\ref{dvanact}) by a set of real chain-coupling
constants $u_k$ in a way which still left the resulting asymmetric
real matrix $H$ tridiagonal,
 \be
 H=
 \left (
 \begin{array}{ccccc}
 \ddots &\ \ \ \ \ \ \ \ \ \ \ddots &&& \\
 \ddots &2+h^2V(x_{-1})&-1-u_0&&\\
 &-1+u_0&2+h^2V(x_{0})&-1-u_1\\
 &&-1+u_1&2+h^2V(x_1)&\ddots\\
  &&& \ddots\ \ \ \ \ \ \ \ \ \ & \ddots
 \ea
 \right )\,.
 \label{trinact}
 \ee
For several special cases of this family of Hamiltonians we then
constructed particular diagonal-matrix metrics $\Theta\neq I$ in
closed form.

This partial success of the project encouraged us to re-open the
question of a {\em complete} solvability of a ${\cal PT}-$symmetric
model in our subsequent paper \cite{fund}. In the cryptohermitian
scenario characterized by the nontriviality of the metrics
$\Theta\neq I$ we found a complete set of metrics for a special case
of Eq.~(\ref{trinact}) with $V(x_k)=0$ (i.e., without any complex
local force) and  with $u_{1}=u_{2}=\ldots = 0$ and
$u_{-1}=u_{-2}=\ldots = 0$, i.e., with the single off-diagonal real
coupling constant ${g}=u_{0}\neq 0$ representing a nonlocal,
manifestly non-Hermitian potential.

In spite of the feasibility of such a construction we still felt
disappointed not only by the necessity of the really lengthy
calculations but also by the comparatively complicated structure of
matrix elements of metrics $\Theta$. Although these elements were
expressible in terms of closed-form polynomials in ${g}$, the degree
of these polynomials {\em grew quickly} with the cut-off dimension
$N$ of the ${g}-$dependent Hilbert space ${\cal H}^{(S)}$. In this
sense  the results of Ref.~\cite{fund} proved discouraging,
especially in the light of a really extreme simplicity of the
underlying single-center interaction.

In our present paper we intend to report a return to optimism.
Firstly, in a preparatory Sec.~\ref{dvedeci} we shall introduce a
double-center model and show that its more complicated dynamics does
not worsen the feasibility of calculations nor a guarantee of the
reality of the energy spectra.  Next, we shall formulate our project
of {\em construction of all the eligible metrics} for this model in
Sec.~\ref{trideci}. In contrast to the similar results of
Ref.~\cite{fund} we shall be able to show here that the present
version of the ${\cal PT}-$symmetric discrete square-well model is
much more friendly since the related menu of metrics $\Theta$
exhibits a paradoxical decrease of complexity during the increase of
dimension $N$. In Sec.~\ref{malodia} (dealing with exceptional, one-
or two-diagonal metrics) and Sec.~\ref{foufideci} (describing all
the remaining metrics) this fact will enable us to find and prove
results valid at all $N$. A compact rigorous proof will be delivered
confirming the validity of our closed and elementary explicit
formulae for metrics $\Theta^{(N)}({\lambda})$ at any given
dimension $N$ and coupling ${\lambda}\in (-1,1)$.

\section{Discrete non-Hermitian square wells\label{dvedeci} }

The key motivation of the present paper resulted from our study of
Refs.~\cite{albev} and \cite{david} where the differential version
(\ref{SEloc}) of a square-well Schr\"{o}dinger equation has been
analyzed with a ${\cal PT}-$symmetric point interaction localized in
the origin (i.e., at $x=0$) and at the two distant points  ($x=\pm
L$), respectively. The former, simpler, single-center arrangement
found its discrete RK analogue in the models of Ref.~\cite{fund}.
Our present paper will offer the discrete RK complement to  the
latter, more ambitious study~\cite{david}.

\subsection{Hamiltonians}

We shall analyze the two-center boundary-interaction $N-$dimensional
models $H^{(N)}({\lambda})$ which form the family
 $$
H^{(3)}({\lambda})=\left[ \begin {array}{ccc}
2&-1-{{\lambda}}&0\\{}-1+{{\lambda}}&2&-1+{{\lambda}}\\{}0&-1-{{\lambda}}&2\end
{array} \right]\,,
 $$
 $$
 H^{(4)}({\lambda})=\left[ \begin {array}{cccc} 2&-1-{\it
{\lambda}}&0&0\\{}-1+{ \it {\lambda}}&2&-1&0\\{}0&-1&2&-1+{\it
{\lambda}}
\\{}0&0&-1-{\it {\lambda}}&2\end {array} \right]\,
 $$
 $$
  H^{(5)}({\lambda})=\left[ \begin {array}{ccccc} 2&-1-{\it
{\lambda}}&0&0&0\\{}-1+{\it {\lambda}}&2&-1&0&0\\{}0&-1&2&-1&0\\{}0
&0&-1&2&-1+{\it {\lambda}}\\{}0&0&0&-1-{\it {\lambda}}&2\end {array}
 \right]
 $$
(etc) and which have the following tridiagonal $N$ by $N$ matrix
form in general,
 \be
  H^{(N)}({\lambda})=  \left[ \begin {array}{cccccc}
 2&-1-{\it {\lambda}}&0&\ldots&0&0
\\
{}-1+{\it {\lambda}}&2&-1&0&\ldots&0
\\
{}0&-1&\ \ \ 2\ \ \ &\ddots&\ddots&\vdots
\\
{}\vdots&0&\ddots&\ \ \ \ddots\ \ \ &-1&0
\\
{}0&\vdots&\ddots&-1&2&- 1+{\it {\lambda}}
\\
{}0&0&\ldots&0&-1-{\it {\lambda}}&2
\end {array}
 \right]\,.
 \label{toym}
 \ee
We intend to pay attention to these models in bound-state regime.
Hence, we have to clarify, first of all, the structure of the domain
of couplings ${\lambda}$ for which the energy spectrum remains real.

\subsection{The reality of the spectra of energies}
%
%
\begin{figure}[h]                     
\begin{center}                         
\epsfig{file=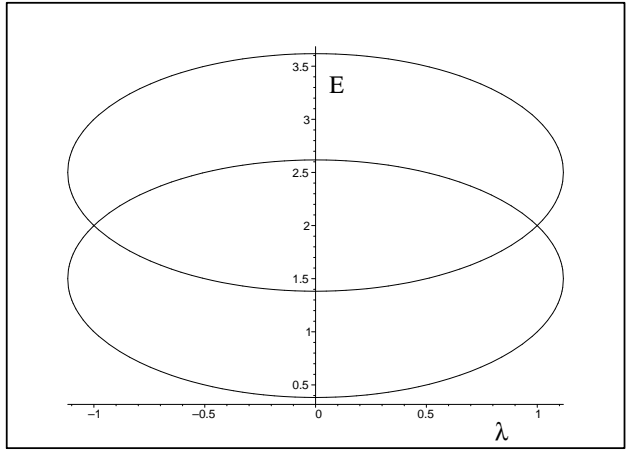,angle=270,width=0.6\textwidth}
\end{center}                         
\vspace{-2mm} \caption{Real spectrum and its degeneracy and
complexification at $N=4$.
 \label{firmonej}}
\end{figure}
%
In the first step, at $N=3$ we obtain the three easily evaluated
eigenvalues
 $$E_0=2\,,\ \ \ \ \  E_{\pm 1}= 2\pm (2-2\,{{\lambda}}^2)^{1/2}\,$$
which are all real and non-degenerate inside the interval of
${{\lambda}}\in (-1,1)$. Next, at $N=4$ we arrive at the four
eigenvalues
 $$  E_{\pm 1/2}=3/2\pm 1/2\,(5-4\,{\lambda}^2)^{1/2}\,,\ \ \ \ \ \
 E_{\pm 3/2}=5/2\pm 1/2\,(5-4\,{\lambda}^2)^{1/2}\,$$
which  stay real within a {\em larger} interval of ${{\lambda}}\in
(-\sqrt{5}/2,\sqrt{5}/2)$. This may seem to indicate that the domain
of the admissible values of ${\lambda}$ may vary with the dimension.
Fortunately, it is not so. Figure \ref{firmonej} clarifies the
apparent puzzle by showing that the condition of non-degeneracy gets
violated precisely at the boundary of the same,
dimension-independent interval ${\lambda}\in (-1,1)$.

%
\begin{figure}[h]                     
\begin{center}                         
\epsfig{file=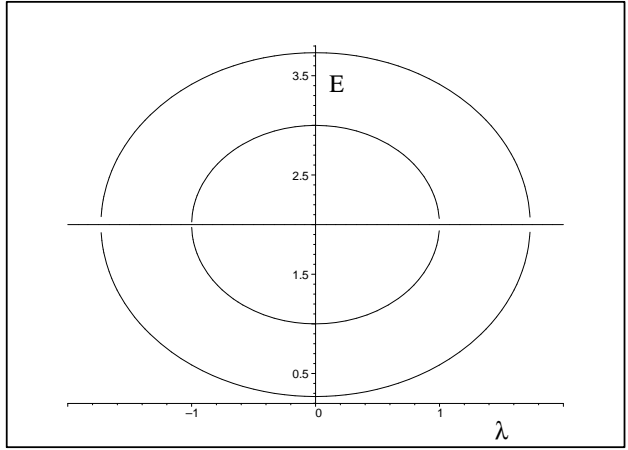,angle=270,width=0.6\textwidth}
\end{center}                         
\vspace{-2mm} \caption{The $\lambda-$dependence of energies at
$N=5$.
 \label{fissrmone}}
\end{figure}

Proceeding to $N=5$ we reveal that the five eigenvalues
 $$
 E_0=2\,,\ \ \ \ \  E_{\pm 1}= 2\pm (1-{\lambda}^2)^{1/2}
 \,,\ \ \ \ \  E_{\pm 2}= 2\pm (3-{\lambda}^2)^{1/2}
 $$
are all real and non-degenerate inside the same interval of
${{\lambda}}\in (-1,1)$. Our  Figure \ref{fissrmone} displays the
coupling-dependence of these energies in full detail.

%
\begin{figure}[h]                     
\begin{center}                         
\epsfig{file=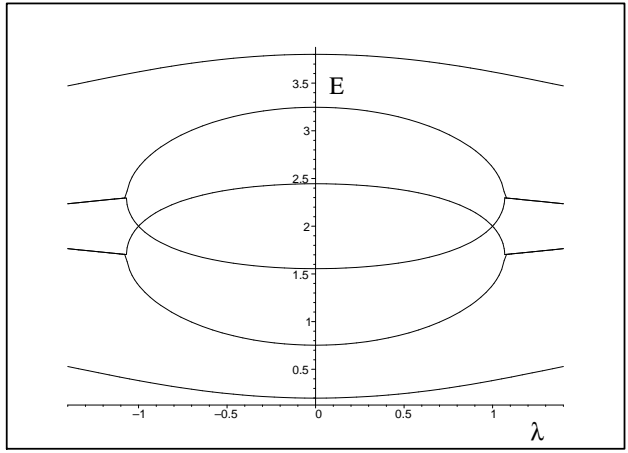,angle=270,width=0.6\textwidth}
\end{center}                         
\vspace{-2mm} \caption{Real parts of energies at $N=6$.
 \label{fisttsrmone}}
\end{figure}
%
In the next case, at $N=6$, the closed formulae for the energies
become clumsy. Still, Figure \ref{fisttsrmone} demonstrates clearly
that the six eigenvalues behave  in  expected manned at all
${\lambda} \in (-1,1)$. At $N=7$ (cf. Figure \ref{fiststsrmone}) the
formulae become, paradoxically, simpler, defining all the spectrum
by the compact equations
 $$
E_0=2 \,,\ \ \ \ \
 E_{\pm 1}=  2\pm 1/2\,\sqrt {8-2\,{{\it {\lambda}}}^{2}-2\,\sqrt {8+{\it {\lambda}}^{4}}
 } \,,\
\ \ \ \  E_{\pm 2}=2\pm \sqrt {2-{{\it {\lambda}}}^{2}}\,,
 $$
 $$\ \ \ \ \
E_{\pm 3}= 2\pm 1/2\,\sqrt {8-2\,{{\it {\lambda}}}^{2}+2\,\sqrt
{8+{{\it {\lambda}}}^{4}}}\,.
  $$
%
%
\begin{figure}[h]                     
\begin{center}                         
\epsfig{file=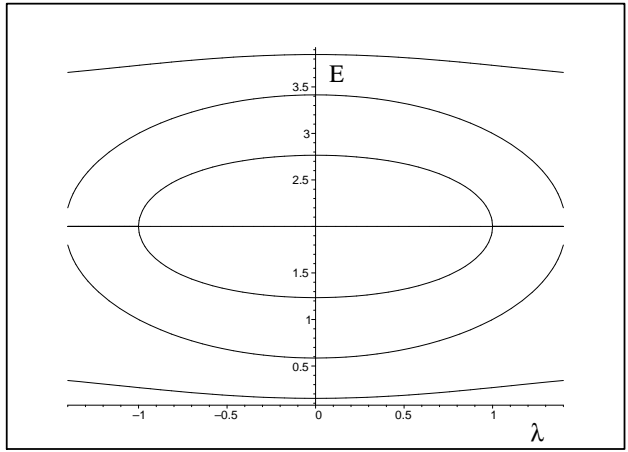,angle=270,width=0.6\textwidth}
\end{center}                         
\vspace{-2mm} \caption{The $\lambda-$dependence of energies at
$N=7$.
 \label{fiststsrmone}}
\end{figure}
%
Next, two separate and rather complicated equations of fourth order
determine the spectrum at even $N=8$ (cf. Figure
\ref{fistutsrmone}). In contrast, our last illustration at $N=9$
(cf. Figure \ref{fisttsrmonez}) yields closed formulae again,
 $$
E_0=2 \,,\ \ \ \ \
 E_{\pm 1}=  2\pm 1/2\,\sqrt {6-2\,{{
\it {\lambda}}}^{2}-2\,\sqrt {{{\it {\lambda}}}^{4}-2\,{{\it
{\lambda}}}^{2}+5}} \,,
 $$
 $$  \ \ \
\ \  E_{\pm 2}=2\pm 1/2\,\sqrt {10-2\,{{ \it
{\lambda}}}^{2}-2\,\sqrt {{{\it {\lambda}}}^{4}+2\,{{\it
{\lambda}}}^{2}+5}}\,,
 $$
 $$\ \ \ \ \
E_{\pm 3}=2\pm 1/2\,\sqrt {6-2\,{{ \it {\lambda}}}^{2}+2\,\sqrt
{{{\it {\lambda}}}^{4}-2\,{{\it {\lambda}}}^{2}+5}}
 \,,
  $$
  $$\
\ \ \ \  E_{\pm 4}=2\pm 1/2\,\sqrt {10-2\,{{ \it
{\lambda}}}^{2}+2\,\sqrt {{{\it {\lambda}}}^{4}+2\,{{\it
{\lambda}}}^{2}+5}}\,.
  $$
%
\begin{figure}[h]                     
\begin{center}                         
\epsfig{file=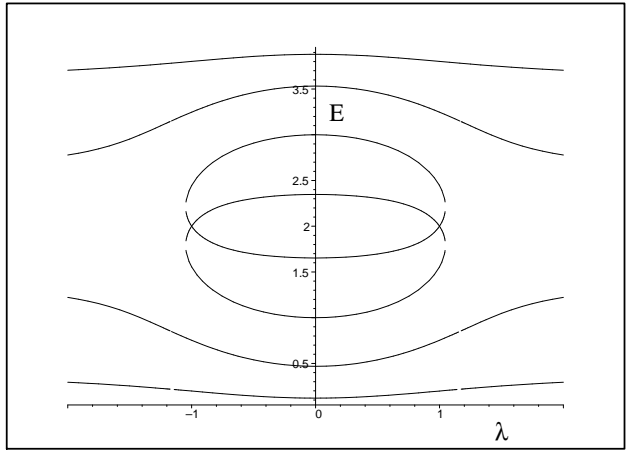,angle=270,width=0.6\textwidth}
\end{center}                         
\vspace{-2mm} \caption{The $\lambda-$dependence of energies at
$N=8$.
 \label{fistutsrmone}}
\end{figure}
%
%
This indicates that the secular polynomials are simpler at odd
dimensions. In all the Figures \ref{firmonej} - \ref{fisttsrmonez}
the interval of the allowed couplings ${\lambda}\in (-1,1)$ does not
vary with the growth of the dimension $N$. An independent
confirmation of this feature of our model will follow, later, from
the existence and invertibility conditions imposed upon the metric
$\Theta$ at any $N$.

%
\begin{figure}[h]                     
\begin{center}                         
\epsfig{file=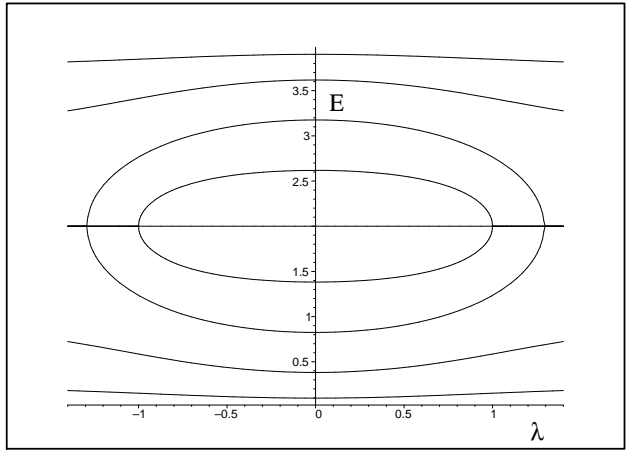,angle=270,width=0.6\textwidth}
\end{center}                         
\vspace{-2mm} \caption{The $\lambda-$dependence of energies at
$N=9$.
 \label{fisttsrmonez}}
\end{figure}
%

Our pictures illustrate that the extreme values of ${\lambda}=\pm 1$
correspond to the triple confluence of the energies $E_{\pm 1}$ with
$E_0$ at odd $N$ and, in our notation, to the incidental degeneracy
of  $E_{+ 1/2}$ with $E_{-3/2}$ at even $N$. In the language of
Refs.~\cite{fragile} one can conclude that at any dimension $N\geq
4$ our model possesses strictly four fragile energies which
complexify during the transition from the Hermitian limit
${\lambda}=0$ to the asymptotic, strongly non-Hermitian regime under
very large $|{\lambda}|\gg 1$.


\section{Hermitization \label{trideci} }

The study of non-Hermitian Hamiltonians $H$ with real spectra is
easier whenever there exists an invertible map $\Omega$  of $H$ upon
an isospectral operator (say, upon $\mathfrak{h}=
\Omega\,H\,\Omega^{-1}$) which is Hermitian (it is, sometimes,
called Dyson's map). The latter operator $\mathfrak{h}$ (acting in
some abstract physical Hilbert space ${\cal H}^{(P)}$, cf.
\cite{Carl2,Geyer,Dieudonne}) may be assumed prohibitively
complicated (otherwise, there would exist no good reason for
studying $H$). The space ${\cal H}^{(P)}$ may be assumed endowed
with the usual Dirac's trivial metric $\Theta^{(P)}=I$. By
definition, the latter space must be unitarily equivalent to its
equally acceptable physical alternative ${\cal H}^{(S)}$
\cite{SIGMA}.

In applications, the role of the Dyson map $\Omega$ degenerates to
the introduction of the elementary product $\Theta=\Omega^\dagger
\Omega$ called metric in ${\cal H}^{(S)}$. For this reason we shall
only be interested here in the constructions of the matrices
$\Theta=\Theta(H)$.

\subsection{The matrices of metrics}

In the theoretical setting outlined in Appendix A below, the key
questions to be answered in connection with the analysis of models
sampled by Eq.~(\ref{toym}) result from the ambiguity of the
assignment of the metric $\Theta$ to a given Hamiltonian $H$
\cite{cubic,Geyer}. For the purposes of clarification of the roots
and forms of this ambiguity the Runge-Kutta discretized Hamiltonians
prove particularly suitable. Indeed, in the related
finite-dimensional (i.e., $N-$dimensional) standard physical Hilbert
spaces ${\cal H}^{(S)}$ the admissible metrics $\Theta=\Theta(H)$ do
only form a finite, strictly $N-$parametric family. Thus, we may
decompose
 \be
   \Theta^{(N)}=
   \sum_{k=1}^N\, {\mu_k}\,{\cal P}_k^{(N)}\,
 \label{e777}
   \ee
and require that the individual Hermitian components ${\cal
P}_k^{(N)}$ of our metric  are some extremely simple pseudometric
(i.e., not necessarily positive definite) matrices with, presumably,
sparse-matrix structure.

From the purely algebraic point of view the correspondence between
$H$ and $\Theta(H)$ is exclusively specified by
Eq.~(\ref{quasihermi}) of Appendix A which can be rewritten in the
explicit linear algebraic form
 \be
 \sum_{k=1}^N\,
 \left [
      \left (H^\dagger\right )_{jk}\,\Theta_{kn}
      -\Theta_{jk}\,H_{kn}\right ] =0
 \,,\ \ \ \ \ j,n=1,2,\ldots,N
   \,.
 \label{htot}
 \ee
At the first sight, the direct use of such a systems of $N^2$
equations for the determination of the matrix elements of
$\Theta=\Theta(H)$ looks discouragingly difficult. Fortunately, not
all of these equations are linearly independent. The number of
unknowns is also lowered by the necessary Hermiticity of acceptable
matrices $\Theta=\Theta^\dagger$. Still, for any given Hamiltonian
$H$, an encouragement and insight into the generic structure of the
solutions $\Theta=\Theta(H)$ may only be acquired step by step, by
the patient solution of Eq.~(\ref{htot}) starting from the smallest
dimensions $N$.

Our first nontrivial real and asymmetric, i.e., ${\cal
PT}-$symmetric and non-Hermitian Hamiltonian
 $
H^{(3)}({\lambda})
 $
has already been studied in the different context (viz., in
connection with the cryptounitary description of scattering, cf.
section 2.1 of Ref.~\cite{discrete}). In the present, bound-state
version of this model we may expect that any eligible  $
H^{(3)}({\lambda})-$dependent metric $\Theta^{(3)}({\lambda})$
compatible with Eq.~(\ref{htot}) will have the real and symmetric
six-parametric matrix form
 $$
\Theta=\left[ \begin {array}{ccc} a&b&c\\{}b&f&g\\{}c&g&m\end
{array} \right]\,.
 $$
The values of its six free real parameters $ a - m$ are only
restricted by the sequence of nine linear relations~(\ref{htot}) and
by the positivity requirement $\Theta>0$. It is easy to verify that
the six nontrivial items of Eq.~(\ref{htot})  number 2, 3, 4, 6, 7
and 8 read, respectively,
                    $ -f + f {{\lambda}} + a + a {{\lambda}} + c + c {{\lambda}}=0$,
                         $ -g + g {{\lambda}} + b - b {{\lambda}}=0$,
                     $-a - a {{\lambda}} - c - c {{\lambda}} + f - f {{\lambda}}=0$,
                    $ -c - c {{\lambda}} - m - m {{\lambda}} + f - f {{\lambda}}=0$,
                         $ -b + b {{\lambda}} + g - g {{\lambda}}=0$ and
                     $-f + f {{\lambda}} + c + c {{\lambda}} + m + m {{\lambda}}=0$.
At ${\lambda}\in (-1,1)$, both the items number 3 and 7 give $b=g$
while the comparison of items number 2 and 8 gives $m=a$. The
remaining four nontrivial equations all coincide with the constraint
$ f(1- {{\lambda}}) = (a + c)(1+ {{\lambda}}) $. We may summarize
that the complete solution of Eq.~(\ref{htot}) has the
three-parametric form
 \be
 \Theta^{(3)}_{(a,b,c)}({\lambda})=
 \left[ \begin {array}{ccc} a&b&c\\{}b&{\frac {
 \left( a+c \right)  \left( 1+{{\lambda}} \right) }{1-{{\lambda}}}}&b\\{}c
&b&a\end {array} \right]\,.
 \label{dian3}
 \ee
At $c=0$ we get the tridiagonal matrix
 $$
 \Theta^{(3)}_{(a,b,0)}({\lambda})=
 \left[ \begin {array}{ccc} a&b&0\\{}b&{\frac {
  a  \left( 1+{{\lambda}} \right)
 }{1-{{\lambda}}}}&b\\{}0
&b&a\end {array} \right]
 $$
which becomes strictly diagonal at $b=0$ where one could also set,
without any loss of generality,
$a=\alpha=(1-{\lambda})/(1+{\lambda})$ yielding
 \be
 \Theta^{(3)}_{(\alpha,0,0)}({\lambda})=
 \left[ \begin {array}{ccc} \alpha &0&0\\{}0&1&0\\{}0
&0&\alpha \end {array} \right]\,,
 \ \ \ \ \ \alpha = {\frac {
    1-{{\lambda}}
 }{1+{{\lambda}}}}\,.
 \label{dia3}
 \ee
At any nonvanishing $\lambda\in (-1,1)$ the latter, diagonal metric
is safely positive and manifestly non-Dirac, $\Theta\neq I$. At the
same time, the positivity of its nondiagonal predecessors must be
guaranteed by an explicit (and not too easy) specification of the
admissible domain of parameters $a$, $b$ and $c$. Without this
guarantee we may only speak about potentially non-invertible or
indefinite pseudometrics replacing, accordingly, also the symbol
$\Theta^{(N)}_{(a,b,c,\ldots)}({\lambda})$ for the metric, say, by a
less specific symbol ${\cal Q}^{(N)}_{(a,b,c,\ldots)}({\lambda})$
whenever appropriate (cf., e.g., Appendix B).



\subsection{The set of simplified pseudometrics}

We shall see below that the feasibility of transition to higher
dimensions $N>3$ will be rendered possible by the Runge-Kutta
algebraization of the Hamiltonian {\em as well as} by our
localization of interaction far from the origin. We shall reveal
that these features of our model open the path toward an enhancement
of efficiency of the construction of $\Theta$ via
expansion~(\ref{e777}) where the individual pseudometric components
${\cal P}$ may be sought in certain simplified sparse-matrix forms.
Naturally, Eq.~(\ref{htot}) may be required to be satisfied also by
every individual component matrix ${\cal P}={\cal P}_k^{(N)}$,
 \be
 \sum_{i=1}^N\,
 \left [
      \left (H^\dagger\right )_{ji}\,{\cal P}_{in}
      -{\cal P}_{ji}\,H_{in}\right ] =0
 \,,\ \ \ \ \ j,n=1,2,\ldots,N
   \,.
 \label{htotbe}
 \ee
The explicit solution of the latter, simplified system of equations
will be further facilitated by the tridiagonal matrix structure of
our Hamiltonians. This can be illustrated at $N=4$ for which the
real, ${\cal PT}-$symmetric and non-Hermitian Hamiltonian
$H^{(4)}({\lambda})$ admits the following real and symmetric ansatz
for the metric
 \be
\Theta^{(4)}({\lambda})= \Theta^{(4)}_{(a,b,c,d)}({\lambda})= \left[
\begin {array}{cccc} a&b&c&d\\{}b&f&g&h
\\{}c&g&m&n\\{}d&h&n&j\end {array}
 \right]\,.
 \label{ctyrak}
 \ee
Out of the sixteen linear relations in (\ref{htot}) only the ones
with numbers 1, 6, 11 and 16 are trivially satisfied. Further, items
3 and 13 give $h=c$, items 7 and 10 yield $m=f$ while the comparison
of 2 with 15 gives $j=a$ and the comparison of 8 with 9 yields
$n=b$. The rest of the set of constraints degenerates to the doublet
of requirements
 $
 f= {\frac {c+a \left( 1+{\it {\lambda}} \right) }{1-{\it {\lambda}}}}
 $
and
 $
 g= {\frac {b+d \left( 1+{\it {\lambda}} \right) }{1-{\it {\lambda}}}}
 $ so that the
exhaustive solution of Eq.~(\ref{htot}) has the four-parametric form
as it should,
 \be
 \Theta^{(4)}_{(a,b,c,d)}({\lambda})=
   \left[ \begin {array}{cccc} a&b&c&d\\{}b&{\frac {c+a
 \left( 1+{\it {\lambda}} \right) }{1-{\it {\lambda}}}}&{\frac {b+d \left( 1+{\it {\lambda}
} \right) }{1-{\it {\lambda}}}}&c\\{}c&{\frac {b+d \left( 1+{ \it
{\lambda}} \right) }{1-{\it {\lambda}}}}&{\frac {c+a \left( 1+{\it
{\lambda}} \right) } {1-{\it {\lambda}}}}&b\\{}d&c&b&a\end {array}
\right].
 \label{dian4}
 \ee
At $d=0$ this matrix becomes pentadiagonal while the additional
constraint $c=0$ makes it tridiagonal. Finally, at $b=0$ we get the
diagonal metric
 \be
 \Theta^{(4)}_{(\alpha,0,0,0)}({\lambda})=
 \left[ \begin {array}{cccc} \alpha
 &0&0&0\\{}0&1&0&0\\{}0&0&1 &0\\{}0&0&0&\alpha
 \end {array} \right]\,,\ \ \ \ \
  \alpha=\frac{1-{\lambda}}{1+{\lambda}}\,.
  \label{dia4}
 \ee
Up to two elements the latter metric coincides with the most common
Dirac's $\Theta^{(Dirac)}=I$. One can expect (and we shall verify
below) that the existence of the similar diagonal metric is a
generic feature of our model at all the dimensions $N=3,4,\ldots$.


\section{Exceptional metrics  \label{malodia}}

The well known ambiguity of the correspondence between Hamiltonian
$H$ and metric $\Theta(H)$ \cite{Geyer} seems to be particularly
important when one tries to improve the precision of RK
approximation and to guess the form of metric $\Theta^{(N+1)}$ from
the knowledge of $\Theta^{(N)}$. For such an extrapolation the
complete solution of Eq.~(\ref{htot}) at the smallest dimensions
$N\leq 4$ does not suffice. Moreover, just a marginal help may be
extracted from the related mathematical literature (dating back to
the early sixties \cite{Dieudonne}). More insight has only been
obtained during the first use of the metrics $\Theta \neq I$ in
nuclear physics \cite{plb,Geyer} and during the development of
${\cal PT}-$symmetric quantum mechanics. In the latter setting an
additional constraint has been accepted and the metric $\Theta$ has
been assumed factorized into the product of parity (i.e., of one of
the pseudometrics ${\cal P}$) with the so called quasiparity
\cite{pseudo} or charge \cite{BBJ}.

Even under the latter class of additional assumptions the metric
$\Theta(H)$ may remain non-unique \cite{cubic,plb,Geyer,Kleefeld}.
This means that the usual choice of the Hamiltonian $H$ should be
accompanied by the additional phenomenological or pragmatic
considerations, i.e., by the physics-dictated or comfort-dictated
specification of some optimal or exceptional hermitizing
metrics~$\Theta(H)$.

\subsection{The strictly diagonal metrics \label{diagonalne} }

In section \ref{trideci} three important generic features of our
present square-well  model have been revealed. Firstly, the eligible
matrices $\Theta$ were found to possess more symmetries than
expected. Secondly, the structure of these matrices appeared to
simplify when one keeps just single first-line matrix element
different from zero. Thirdly, the simplest form of matrix elements
seems to be achieved when the first-line matrix elements $a, b,
\ldots$ are properly rescaled in a coupling-dependent way.

In practice, the most prominent role will always be played by the
matrices $\Theta\neq I$ which are as close to diagonal ones as
possible having $a\,(=\Theta_{11})\neq 0$ while
$b\,(=\Theta_{12})=c\, (=\Theta_{13})=\ldots=0$. In this sense our
first present interesting result is the following one.

\begin{thm}
For any matrix Hamiltonian (\ref{toym}) with coupling $\lambda \in
(-1,1)$ and dimension $N=3,4,\ldots$ there always exists a diagonal,
positive definite metric matrix which differs from the Dirac's
$\Theta=I$ just at the elements $\Theta_{kk}$ with $k=1$ and $k=N$.
Its explicit form is
 \be
 \Theta^{(N)}_{(\alpha,0,\ldots,0)}({\lambda})=
  \left[ \begin {array}{ccccc}
 \alpha &0&\ldots&0&0
 \\{}0&1&0&\ldots&0\\
 {}\vdots&\ddots&\ddots&\ddots&\vdots
 \\{}0&\ldots&0&1&0
 \\{}0&0&\ldots&0&\alpha
 \end {array} \right]\,,\ \ \ \ \alpha=\alpha(\lambda)={\frac
 {1-{\it {\lambda}}}{1+{\it {\lambda}}}}\,.
 \label{diago}
 \ee

\end{thm}

 \bp
Equation (\ref{diago}) coincides with Eq.~(\ref{dia3}) at $N=3$ and
with Eq.~(\ref{dia4}) at $N=4$. At any integer $N>4$ the insertion
of ansatz (\ref{diago}) converts Eq.~(\ref{htot}) into  identity.
 \ep

\begin{coro}
The spectrum of Hamiltonian (\ref{toym}) with coupling $\lambda \in
(-1,1)$ is real at any dimension $N=3,4,\ldots$.
\end{coro}

 \bp
 The existence of the metric guarantees that in the corresponding
 finite-dimensional Hilbert space
 the Hamiltonian matrix is Hermitian.
 \ep

Just an inessential modification of the above construction leads
also to the following interesting observation.

\begin{prop}
For any matrix Hamiltonian (\ref{toym}) with coupling $\lambda \in
(-1,1)$ and dimension $N=3,4,\ldots$ there always exists an
antidiagonal, indefinite pseudometric  matrix of the form
 \be
 {\cal Q}^{(N)}_{(0,0,\ldots,0,\alpha)}({\lambda})=
  \left[ \begin {array}{ccccc}
 0&0&\ldots&0&\alpha
 \\{}0&\ldots&0&1&0\\
 {}\vdots&
 {\large \bf _. } \cdot {\large \bf ^{^.}}&
 {\large \bf _. } \cdot {\large \bf ^{^.}}
 &
 {\large \bf _. } \cdot {\large \bf ^{^.}}&\vdots
 \\{}0&1&0&\ldots&0
 \\{}\alpha&0&\ldots&0&0
 \end {array} \right]\,,\ \ \ \ \alpha=\alpha(\lambda)={\frac
 {1-{\it {\lambda}}}{1+{\it {\lambda}}}}\,.
 \label{andiago}
 \ee

\end{prop}

 \bp
At $N=3$, equation (\ref{andiago}) degenerates to formula
(\ref{dian3}) at $a=b=0$ and $c=\alpha$. At $N=4$, equation
(\ref{andiago}) degenerates to formula (\ref{dian4}) at $a=b=c=0$
and $d=\alpha$. At any higher integer $N>4$ the insertion of ansatz
(\ref{andiago}) converts Eq.~(\ref{htot}) into  identity. At the
same time one can easily verify that matrix (\ref{andiago}) is not
positive definite at any $N \geq 2$.
 \ep

\begin{pozn}
It is worth noticing that {\em before} one performs the continuous
$h \to 0$ limit of the present Runge-Kutta picture, certain metric
operators obtained in Ref.~\cite{cubic} at  $\lambda=0$ may be
interpreted as positive definite superpositions of our two
one-diagonal matrices (\ref{diago}) and (\ref{andiago}).
\end{pozn}

\subsection{The strictly bidiagonal pseudometrics \label{bidiagonalne} }

In our present concrete realization of expansion (\ref{e777}) the
general matrix of metric $\Theta^{(N)}_{(a,b,\ldots)}({\lambda})$
will always be written as a linear superposition of the following
$N-$plet of specific pseudometric matrices
 \ben
 {\cal P}_1^{(N)}({\lambda})\sim {\cal Q}^{(N)}_{(a,0,\ldots,0)}({\lambda})\,,\ \ \
 {\cal P}_2^{(N)}({\lambda})\sim {\cal Q}^{(N)}_{(0,b,0,\ldots,0)}({\lambda})\,,\ \
 \ldots\,.
 \een
Their mutual independence is trivially guaranteed by the presence of
the mere single nonvanishing matrix element in their respective
first lines.


On the level of our above-mentioned $N\leq 4$ experience and
constructions we just managed to guess and prove the generic form of
the diagonal and antidiagonal pseudometrics ${\cal P}_1({\lambda})$
[cf. Eq.~(\ref{diago})] and ${\cal P}_N({\lambda})$ [cf.
Eq.~(\ref{andiago})]. Further insight must be built using the next,
$N=5$ ansatz for the real and symmetric metric $\Theta^{(5)}$. Using
symbolic manipulations we managed to reveal that without any loss of
generality this ansatz may be written in the simplified, symmetrized
nine-parametric form
 \be
\Theta^{(5)}_{(a,b,c,d,e)}= \left[ \begin {array}{ccccc}
a&b&c&d&{\it e}\\{}b&f&g&h&{\it d}\\{}c& g&m&g&c\\{}d&h&g&f&b
\\{} {\it e}&{\it d}&c&b&a\end {array} \right]\,.
\label{petorka}
 \ee
Out of the twenty five independent relations (\ref{htot}) we managed
to eliminate nine trivial ones yielding the four quadruplets of
equations which enabled us to eliminate $m=f+h-c(1+{\lambda})$ (from
items 8, 12, 14 and 18), $g=(b+d)/(1-{\lambda})$ (from items 3, 11,
15 and 23), $ h= (c+e(1+{\lambda}))/(1-{\lambda}) $ (from items  4,
10, 16 and 22) and $ f= (c+a(1+{\lambda}))/(1-{\lambda}) $ from
items 2, 6, 20 and 24. Thus, we arrived at the complete
five-parametric solution $\Theta^{(5)}_{(a,b,c,d,e)}({\lambda})$ of
Eq.~(\ref{htot}),
 $$
    \left[ \begin {array}{ccccc} a&b&c&d&{\it e}\\{}b&{
\frac {c+a \left( 1+{\it {\lambda}} \right) }{1-{\it
{\lambda}}}}&{\frac {b+d}{1-{ \it {\lambda}}}}&{\frac {c+{\it e}\,
\left( 1+{\it {\lambda}} \right) }{1-{\it {\lambda}}}
}&d\\{}c&{\frac {b+d}{1-{\it {\lambda}}}}&{\frac {c+a \left( 1+{\it
{\lambda}} \right) }{1-{\it {\lambda}}}}+{\frac {c+{\it e}\, \left(
1+{\it {\lambda}} \right) }{1-{\it {\lambda}}}}-c \left( 1+{\it
{\lambda}} \right) &{\frac {b+d}{1 -{\it {\lambda}}}}&c\\{}d&{\frac
{c+{\it e}\, \left( 1+{\it {\lambda}} \right) }{1-{\it
{\lambda}}}}&{\frac {b+d}{1-{\it {\lambda}}}}&{\frac {c+a
 \left( 1+{\it {\lambda}} \right) }{1-{\it {\lambda}}}}&b\\{}{\it e
}&d&c&b&a\end {array} \right]\,.
 $$
At $e=0$ this matrix is seven-diagonal. After we set $d=0$ it
becomes pentadiagonal while the next constraint $c=0$ makes it
tridiagonal. The last, $b=0$ item in the list of simplifications is
the diagonal matrix which is compatible with Eq.~(\ref{diago}) and
non-Dirac, $\Theta^{(5)}_{(a,0,0,0,0)}({\lambda})\neq I$ at $\lambda
\neq 0$.

The most important consequence of our  ``brute-force" calculations
performed at $N=5$ lies in the highly desired clarification of
certain higher$-N$ tendencies exhibited by nondiagonal metrics. In
the light of this experience we were able to formulate and prove the
following

\begin{prop}
At any dimension $N=3,4,\ldots$ and coupling $\lambda \in (-1,1)$,
pseudo-Hermiticity relation (\ref{htotbe}) is satisfied by
Hamiltonian (\ref{toym}) and by the bidiagonal pseudometric matrix
 \be
 {\cal P}^{(N)}_2({\lambda})=
 {\cal Q}^{(N)}_{(0,\beta,0,0,\ldots,0)}({\lambda})=
 \left[ \begin {array}{cccccc}
  0&\beta
 &0&0&\ldots&0\\{}
 \beta&0
 &1&0&\ldots&0\\{}0&1&\ddots&\ddots&\ddots&\vdots
 \\{}0&\ddots&\ddots&0&1&0
 \\{}\vdots&\ddots&0&1&0 &\beta
 \\{}0&\ldots&0&0&\beta&0
 \end {array} \right]\,
 \label{bidia}
 \ee
containing two pairs of $\lambda-$dependent elements
$\beta=\beta(\lambda)={1-{\it {\lambda}}}$ connected by two unit
diagonals.
\end{prop}
 \bp
  Equation (\ref{bidia}) is compatible with Eq.~(\ref{dian3}) at
$N=3$ and with Eq.~(\ref{dian4}) at $N=4$. At any integer $N>4$ the
insertion of ansatz (\ref{bidia}) converts Eq.~(\ref{htotbe}) into
identity.
 \ep

\begin{prop}
At any dimension $N=3,4,\ldots$ and coupling $\lambda \in (-1,1)$,
Hamiltonian (\ref{toym}) and pseudometric matrix
 \be
 {\cal P}^{(N)}_{N-1}({\lambda})=
 {\cal Q}^{(N)}_{(0,0,\ldots,0,\beta,0)}({\lambda})=
 \left[ \begin {array}{cccccc}
                     0&\ldots&0&0&\beta&0\\{}
                     0&\ldots&0&1&0&\beta\\{}
                     \vdots
                     &{\large \bf _. } \cdot {\large \bf ^{^.}}
                     &{\large \bf _. } \cdot {\large \bf ^{^.}}
                     &0
                     &1&0
 \\{}0&1&{\large \bf _. } \cdot {\large \bf ^{^.}}
 &{\large \bf _. } \cdot {\large \bf ^{^.}}
 &{\large \bf _. } \cdot {\large \bf ^{^.}}
 &\vdots
 \\{}\beta&0&1&0&\ldots&0
 \\{}0&\beta&0&0&\ldots&0
 \end {array} \right]\,
 \label{anbidia}
 \ee
with two antidiagonals and four elements $\beta={1-{\it {\lambda}}}$
satisfy the pseudo-Hermiticity relation (\ref{htotbe}).
\end{prop}
 \bp
  Equation (\ref{anbidia}) is compatible with Eq.~(\ref{dian3}) at
$N=3$ and with Eq.~(\ref{dian4}) at $N=4$. At any integer $N>4$ the
insertion of ansatz (\ref{anbidia}) converts Eq.~(\ref{htotbe}) into
identity.
 \ep
We see that at $N>3$ the same simplification of the interior matrix
elements to units or zeros occurs in both the bidiagonal and
antibidiagonal cases. Similar phenomenon will characterize also the
structure of all the remaining elements of our set of pseudometrics
${\cal P}^{(N)}_k={\cal P}^{(N)}_k(\lambda)$.

\section{The complete set of pseudometrics \label{foufideci}}

The simplest pseudometric matrices ${\cal P}^{(N)}_k({\lambda})$
entering expansion (\ref{e777}) were defined by Eq.~(\ref{diago})
(where we have to set $k=1$), by Eq.~(\ref{bidia}) (where $k=2$), by
Eq.~(\ref{andiago}) (where $k=N$), and by Eq.~(\ref{anbidia}) (where
$k=N-1$). On the basis of inspection of metrics $\Theta^{(N)}$
evaluated at dimensions $N \leq 5$ we are now prepared to guess and
prove the explicit form of the remaining pseudometrics  ${\cal
P}^{(N)}_k({\lambda})$ at all the subscripts $k$ such that $3 \leq k
\leq N-2$. Our $N$ by $N$ candidates ${\cal C}^{(N)}_k$ for these
pseudometrics will be sparse matrices with the property
 \be
 \left (
 {\cal C}^{(N)}_k
 \right )_{mn}=0\ \ \ \ {\rm whenever}\ \ \ \ \ m+n\equiv k\ ({\rm
 mod}\
 2)\,.
 \label{single}
 \ee
This means that on a sufficiently large chessboard the nonvanishing
matrix elements of each of these matrices would only occupy either
black or white fields. Secondly, all of these ``colored" (i.e.,
black or white) ansatz matrices will share the following
two-parametric form of their upper segment,
 \be
 {\cal C}^{(N)}_k={\cal C}^{(N)}_k(z,v)=
 \left[ \begin {array}{ccccccc}
                    \ldots&0&0&z&0&0&\ldots
  \\{}\ldots&0&v&0&v&0&\ldots
  \\{}
 {\large \bf _. } \cdot {\large \bf ^{^.}}
 &v&0&1&0&v&\ddots
  \\{}
 {\large \bf _. } \cdot {\large \bf ^{^.}}
 &0&1&0&1&0&\ddots
  \\{}
 {\large \bf _. } \cdot {\large \bf ^{^.}}
 &1&0&1&0&
 1&\ddots
  \\{}
 {\large \bf _. } \cdot {\large \bf ^{^.}}
 &0&1&0&1&0&\ddots
 \\{}
 {\large \bf _. } \cdot {\large \bf ^{^.}}
 &
 {\large \bf _. } \cdot {\large \bf ^{^.}}
 &
 {\large \bf _. } \cdot {\large \bf ^{^.}}&\vdots&\ddots&\ddots&\ddots
\end {array} \right]\,
 \label{endiago}
 \ee
copied and shared also by the  $\pi/2-$rotated  left segment, lower
segment and right segment. In compact notation we are led to the
left-right as well as up-down asymmetric arrays of matrix elements,
 \be
  {\cal C}^{(N)}_k(z,v)=\left[ \begin {array}{ccccccccc}
  0&0&\ldots&0&z&0&0&\ldots&0
 \\{}0&\ldots&0&v&0&v&0&\ddots&\vdots
 \\{}\vdots&
 {\large \bf _. } \cdot {\large \bf ^{^.}}
 &{\large \bf _. } \cdot {\large \bf ^{^.}}&
 {\large \bf _. } \cdot {\large \bf ^{^.}}
 &1&\ddots&\ddots&\ddots&0
 \\{}0&v&{\large \bf _. } \cdot {\large \bf ^{^.}}
 &{\large \bf _. } \cdot {\large \bf ^{^.}}
 &&\ddots&\ddots&v&0
 \\{}z&0&1&& \ddots&&1&0&z
 \\{}0& v&\ddots&\ddots&&
 {\large \bf _. } \cdot {\large \bf ^{^.}}
 &{\large \bf _. } \cdot {\large \bf ^{^.}}&v&0
 \\{}\vdots&\ddots&\ddots&\ddots
 &1&{\large \bf _. } \cdot {\large \bf ^{^.}}
 &{\large \bf _. } \cdot {\large \bf ^{^.}}
 &{\large \bf _. } \cdot {\large \bf ^{^.}}&\vdots
 \\{}0&\ldots&0&v&0&v&0&\ldots&0
 \\{}0&0&\ldots&0&z&0&\ldots&0&0\end {array} \right]\,.
 \ee
In these two-parametric matrices the uppermost item $z$ sits in the
$k-$th place of the first row of our matrix while the bottom line
contains $z$ in the $(N+1-k)-$th place. Together with the other two
$z$'s we obtain the quadruplet of vertices of connected by the four
diagonal rows of elements $v$ forming a parallelogram. Keeping still
in mind the chessboard visualization of matrices ${\cal
C}^{(N)}_k(z,v)$ we require that this parallelogram separates the
outside domain (filled just by zeros) from the rhomboidal interior
filled by ``equal color" units and ``opposite color" zeros. A few
illustrative samples of explicitly computed matrices ${\cal P}$
possessing this structure may be found in Appendix B below. We are
now prepared to prove our final result.

\begin{thm}
 \label{hlavni}
Pseudometric matrices ${\cal P}^{(N)}_k({\lambda})$ with $3 \leq k
\leq N-2$ which would be compatible with Hamiltonian (\ref{toym})
via condition~(\ref{htotbe}) may be identified with the
$\lambda-$dependent matrices ${\cal C}^{(N)}_k(\gamma,\delta)$ where
 \be
  \gamma=\gamma({\lambda})
 ={\frac {1-{\it {\lambda}}}{1+{\it {\lambda}}^2}}
 \,,\ \ \ \
  \delta=\delta({\lambda})={\frac {1}{1+{\it {\lambda}}^2}}\,.
 \label{defi}
  \ee
\end{thm}
 \bp
The inspection of concrete examples using small fixed dimensions $N$
(obtainable by the algorithm outlined  in Appendix B) indicates that
all our elementary pseudometrics ${\cal P}_k({\lambda})$ may be
expected to exhibit the above-mentioned fourfold symmetry. This
expectation is easily verified by immediate insertions confirming
that in our algebraic manipulations with Eq.~(\ref{htotbe}) it is
sufficient to work just with the not too large submatrices
(\ref{endiago}) glued, if necessary, to their rotated neighbors. In
this sense, the role of the position of the subscript $k$ in
interval $[3,N-2]$ remains inessential.

In the next preparatory step of our  proof let us recall the
detailed form of our generic ansatz (\ref{endiago}) and visualize it
also as written on the black-and-white chessboard. Obviously, the
color of fields with nonvanishing matrix elements will be fixed as
``white" or ``black" for each subscript $k$. At the same time, our
tridiagonal Hamiltonian $H$ will be both ``white" (= its main
diagonal, $H^{(w)}:=2\,I$) and ``black" (= its upper diagonal,
$H^{(b+)}({\lambda})$, as well as its lower diagonal,
$H^{(b-)}({\lambda})$). This type of coloring simplifies our
argumentation because in our fundamental Eq.~(\ref{htotbe}) we may
use any ${\cal P}_k({\lambda})={\cal C}_k({\gamma,\delta})$ and
decompose
$H=H({\lambda})=H^{(w)}+H^{(b+)}({\lambda})+H^{(b-)}({\lambda})$ and
$H^\dagger=H(-{\lambda})=H^{(w)}+H^{(b+)}(-{\lambda})+H^{(b-)}(-{\lambda})$.
Obviously, the `` color" of any selected ${\cal
C}_k({\gamma,\delta})$ will be shared by its products with $H^{(w)}$
and it will differ from its products with $H^{(b\pm)}(\pm
{\lambda})$. Using this idea we may very quickly deduce that
starting from Eq.~(\ref{endiago}) we shall {\em always} obtain
product ${\cal C}\,H$ in the form characterized by its upper segment
 $$
  \left[ \begin {array}{ccccccccc}
                    \ldots&0&0&\underline{-\gamma}&2\gamma&\underline{-\gamma}&0&0&\ldots
  \\ &0&-\delta&2\delta&\underline{-2\delta}&2\delta&-\delta&0&
  \\
 {\large \bf _. } \cdot {\large \bf ^{^.}}
 &-\delta&2\delta&-1-\delta&2&-1-\delta&2\delta&-\delta&\ddots
  \\{}
 {\large \bf _. } \cdot {\large \bf ^{^.}}
 &2\delta&-1-\delta&2&-2&2&-1-\delta&2\delta&\ddots
  \\{}
 {\large \bf _. } \cdot {\large \bf ^{^.}}
 &-1-\delta&2&-2&2&-2&2&-1-\delta&\ddots
  \\{}
 {\large \bf _. } \cdot {\large \bf ^{^.}}
 &
 {\large \bf _. } \cdot {\large \bf ^{^.}}
 &
 {\large \bf _. } \cdot {\large \bf ^{^.}}
 &
 {\large \bf _. } \cdot {\large \bf ^{^.}}
 &
 \vdots&\ddots&\ddots&\ddots&\ddots
%
\end {array} \right]\,.
 \label{rightendiago}
 $$
Similarly, the product $H^\dagger\,{\cal C}$ will be specified by
its very similar upper segment
 $$
  \left[ \begin {array}{ccccccc}
                    \ldots&0&
                    \underline{-\delta(1-{\lambda})}&2\gamma&\underline{-\delta(1-{\lambda})}&0&\ldots
  \\ &-\delta&2\delta&\underline{-1-\gamma(1+{\lambda})}&2\delta&-\delta&
  \\
 {\large \bf _. } \cdot {\large \bf ^{^.}}
 &2\delta&-1-\delta&2&-1-\delta&2\delta&\ddots
  \\{}
 {\large \bf _. } \cdot {\large \bf ^{^.}}
 &-1-\delta&2&-2&2&-1-\delta&\ddots
  \\{}
 {\large \bf _. } \cdot {\large \bf ^{^.}}
 &2&-2&2&-2&2&\ddots
  \\{}
 {\large \bf _. } \cdot {\large \bf ^{^.}}
 &
 {\large \bf _. } \cdot {\large \bf ^{^.}}
 &
 {\large \bf _. } \cdot {\large \bf ^{^.}}
 &
 \vdots&\ddots&\ddots&\ddots
%
\end {array} \right]\,.
 \label{leftendiago}
 $$
The difference between these two upper-part matrix structures only
involves the underlined matrix elements. It is easy to check that
the coincidence of these two remaining triplets of underlined matrix
elements is guaranteed since it follows from definition (\ref{defi})
of quantities $\gamma=\gamma({\lambda})$ and
$\delta=\delta({\lambda})$.

The symmetries of our pseudometrics imply that the same coincidence
of matrix elements will take place for the two lower parts of
products ${\cal C}\,H$ and $H^\dagger\,{\cal C}$. In contrast,  the
rotation of the upper-part matrices by the mere $\pm \pi/2$ changes
the picture. Different pattern emerges causing, fortunately, just an
exchange of the underlined matrix elements between the respective
left or right parts of products ${\cal C}\,H$ and $H^\dagger\,{\cal
C}$. In this way we arrive at another, equivalent matrix
representation of equation $H^\dagger\,{\cal C}={\cal C}\,H$.
Indeed, in its new form this representation can be made explicit,
say, via its left-part embodiment
 $$
 \left[ \begin {array}{cccc}
  \vdots&{\large \bf _. } \cdot {\large \bf ^{^.}}
  &{\large \bf _. } \cdot {\large \bf ^{^.}}
  &{\large \bf _. } \cdot {\large \bf ^{^.}}
  \\{}0&0&-\delta&{\large \bf _. } \cdot {\large \bf ^{^.}}
 \\{}0&-\delta&2\,\delta&{\large \bf _. } \cdot {\large \bf ^{^.}}
 \\{}\underline{-\gamma}&2\,\delta&-1-\delta&{\large \bf _. } \cdot {\large \bf ^{^.}}
 \\{}2\,\gamma&\underline{-2\,\delta}&2&\ldots
 \\{}\underline{-\gamma}&2\,\delta&-1-\delta&\ddots
 \\{}0&-\delta&2\,\delta&\ddots
 \\{}0&0&-\delta&\ddots
 \\{}\vdots&\vdots&\vdots&\ddots
 \end {array} \right]
 =\left[ \begin {array}{cccc}
  \vdots&{\large \bf _. } \cdot {\large \bf ^{^.}}
  &{\large \bf _. } \cdot {\large \bf ^{^.}}
  &{\large \bf _. } \cdot {\large \bf ^{^.}}
  \\{}0&0&-\delta&{\large \bf _. } \cdot {\large \bf ^{^.}}
 \\{}0&-\delta&2\,\delta&{\large \bf _. } \cdot {\large \bf ^{^.}}
 \\{}\underline{-\delta(1-{\lambda})}&2\,\delta&-1-\delta&{\large \bf _. } \cdot {\large \bf ^{^.}}
 \\{}2\,\gamma&\underline{-1-\gamma(1+{\lambda})}&2&\ldots
 \\{}\underline{-\delta(1-{\lambda})}&2\,\delta&-1-\delta&\ddots
 \\{}0&-\delta&2\,\delta&\ddots
 \\{}0&0&-\delta&\ddots
 \\{}\vdots&\vdots&\vdots&\ddots
 \end {array} \right]\,.
 $$
We see that the modifications are inessential and that the rotated
recipe leads to the same identities. We may conclude that the
products $H^\dagger\,{\cal C}$ and ${\cal C}\,H$ coincide. This
completes the proof of eligibility of our ansatz (\ref{endiago}) +
(\ref{defi}) for independent non-exceptional sparse-matrix
components ${\cal P}_k^{(N)}(\lambda)$ of the general $N-$parametric
metric at any dimension $N$.
 \ep

\section{Summary and discussion}

We can summarize that for our square-well model (\ref{toym}) the
$N-$term decomposition (\ref{e777}) of {\em all the existing}
metrics $\Theta^{(N)}(\lambda)$ may be written in terms of the
non-exceptional closed-form pseudometrics of Theorem \ref{hlavni}
complemented by the exceptional closed-form metric ${\cal
P}^{(N)}_1({\lambda})$ [defined by Eq.~(\ref{diago})] and
closed-form pseudometrics ${\cal P}^{(N)}_2({\lambda})$ [defined by
Eq.~(\ref{bidia})], ${\cal P}^{(N)}_{N-1}({\lambda})$ [defined by
Eq.~(\ref{anbidia})] and ${\cal P}^{(N)}_N({\lambda})$ [defined by
Eq.~(\ref{andiago})].

In a broader perspective, the existence of remarkable parallels as
well as differences between the standard and ${\cal PT}-$symmetric
models of bound states has been reconfirmed.  It has been emphasized
that in the former case one works, as a rule, solely with the most
elementary Dirac metric $\Theta=I$.  In the majority of ${\cal
PT}-$symmetric models, on the contrary, the technically most
difficult task concerns the construction of the appropriate metric
or metrics $\Theta \neq I$. For this reason, the work with the {\em
difference} (rather than with the more current differential) ${\cal
PT}-$symmetric Schr\"{o}dinger equations has been preferred and may
be recommended as easier in technical terms.

In our present paper we intended to address and formulate the
problem of solvability in ${\cal PT}-$symmetric context. We felt
inspired by one of the simplest differential square-well
Hamiltonians where a ``minimal" non-Hermiticity has been introduced
via boundary conditions and where an exceptionally elementary metric
$\Theta$ has been found in Ref.~\cite{david}. On this background we
selected and analyzed the difference-equation version (\ref{toym})
of this model.

A supplementary reason for our choice of model (\ref{toym}) has been
provided by Refs.~\cite{fragile} and \cite{Jones} where several
physical and dynamical assumptions (say, about a large distance
between interaction centers, etc) have been made in the technically
more difficult differential-equation context. In our study of the
discrete sample (\ref{toym}) of a generic boundary-condition model
some of the empirical observations made in these references
(concerning, e.g., the correspondence between the fragile and robust
energy levels \cite{fragile}) reappeared and have been illustrated
by a few pictures.

The most interesting conclusions may be extracted from the
comparison of our present results with their predecessors described
in Ref.~\cite{fund}. In both these cases point-like interactions
were used. Still, the decisive advantage of their present version
has been found in their consequent localization in the closest
vicinity of the boundaries. Formally, this feature of our $H$ has
been reflected by an enormous simplification of the structure of
{\em all} of the related matrices ${\cal P}$ (= pseudometrics) and
$\Theta$ (= metrics).

The latter merit of our model was quite unexpected and its
explanation also forms a mathematical core of our present message.
{\it A posteriori} we may conclude that the simplification of our
matrix equations for ${\cal P}$ and $\Theta$ resulted from a
``hidden" possibility of their split in the ``white" and ``black"
components. This is also a deeper reason why the simplicity of the
metrics attached to our present model is in a sharp contrast with
the complicated recurrent nature of the analogous matrices in the
older model of Ref.~\cite{fund}.

In connection with this feature of our model the core of feasibility
as well as of the rigorous form of our constructions may be seen in
the availability of appropriate ansatzs. They were found by
extrapolation from investigative constructions performed, at the
smallest dimensions, by the brute-force linear-algebraic techniques.
{\it A posteriori} we must appreciate, therefore, the drastic
reduction of the large set of algebraic quasihermiticity conditions
(\ref{htot}) or (\ref{htotbe}) to the mere double definition
(\ref{defi}) of functions $\gamma({\lambda})$ and
$\delta({\lambda})$. This reduction seems to have been caused by a
certain purely formal interplay between the tridiagonality of $H$
and its free-motion character preserved near the origin. It was
precisely this fine-tuned dynamical input which suppressed the
computational difficulties and which facilitated, decisively, the
explicit interactive and extrapolative analysis of $\Theta(H)$.

Our closed formulae appear transparent, especially if one decides to
work, say, with just a few terms in the general series~(\ref{e777}).
In this way even the very pragmatic users of the discretized norm or
inner product
 \ben
 \langle\!\langle \psi|\phi
 \rangle:=\langle \psi|\phi \rangle^{(S)}=
 \sum_k\,\sum_n\,\psi^*(x_k) \Theta_{k,n}\,\phi(x_n)
 \ \ \ \ {\rm in}
 \ \  {\cal H}^{(S)}\,
 \een
might employ not only the diagonal matrix metric of
Eq.~(\ref{diago}) but also, say, a Sobolev-space resembling
discretized inner product with tridiagonal
 $
   \Theta^{(N)}=
   2\,{\cal P}_1^{(N)}- \gamma\,{\cal P}_2^{(N)}
 $ where, even at large $N$, the obligatory guarantee of positive
 definiteness would just
require that $|\gamma|<1$.

In a more general framework of study of  mutually non-equivalent
possible physical Hilbert spaces ${\cal H}^{(S)}$ assigned to a
given Hamiltonian $H$ we believe that  a deeper insight if not
classification could be obtained in the nearest future, especially
on the ambitious level aiming at the less elementary interactions
and/or more complicated combinations of independent components in
the positive definite matrices of metrics.

\newpage

\section*{Appendix A: ${\cal PT}-$symmetric models}

One of the most common tacit assumptions which lies at the very
heart of the correct physical interpretation of Eqs.~(\ref{SEloc})
and/or~(\ref{SEdis}) is that the real variables $x$ and/or $x_k$
represent an experimentally measurable quantity (most often, a
coordinate or momentum of a point particle). This type of postulate
has been declared redundant in the so called ${\cal PT}-$symmetric
models where all $x$ or $x_k$ are allowed complex (cf. Refs.
\cite{Carl} for a full account of the theory). The endpoints $\pm L$
may be then chosen as any left-right symmetric pair of points,
finite or infinite, in complex plane of $x$.

This is a new freedom. It implies, e.g., that the ``wrong-sign"
potential $V(x) = -x^4$ becomes tractable as a fully legal source of
a discrete and real spectrum of bound-state energies which is
bounded from below. One only has to complexify the points of
boundary in an appropriate left-right symmetric (called, for
historical reasons, ${\cal PT}-$symmetric) manner. In a typical
replacement $+L \to \varrho \exp (-{\rm i} \varphi)$ and $-L \to
-\varrho \exp (+{\rm i} \varphi)$ one uses a very large (or
infinite) real $\varrho \gg 1$ and some safely nonvanishing real
angle $\varphi$ (cf., e.g., Refs.~\cite{BGMateo} for more details).

In the ${\cal PT}-$symmetric scenario the role of the variable $x$
(and, {\it mutatis mutandis}, of $x_k$) is purely auxiliary. In
principle, this variable does not represent an eigenvalue of any
operator of observable even when it remains real. Mathematically one
speaks about a ``false" Hilbert space ${\cal H}^{(F)}$ equipped with
the most common definition of the inner product of wave functions,
 \be
 \br \psi_a|\psi_b\kt\
 \left (=\br \psi_a|\psi_b\kt^{(F)}
 \right )\
 =\int_{-L}^{L}\,\psi_a^*(x)\psi_b(x)\,dx\,
 \label{prod}\ \ \ \ \ {\rm in} \ \ \ \ \ {\cal H}^{(F)}\,.
 \ee
Hamiltonians are represented there, typically, by non-self-adjoint
differential or difference operators,
 \be
 H=-\frac{d^2}{dx^2} + V(x) \neq H^\dagger\ \ \ \ \ {\rm in} \ \ \ \ \ {\cal H}^{(F)}.
 \ee
As long as they have to generate a unitary time evolution, we must
change the definition of Hermitian conjugation in order to make them
properly Hermitian in the resulting ``standard", i.e.,  physical
Hilbert space ${\cal H}^{(S)}$,
 \be
 H=-\frac{d^2}{dx^2} + V(x) = H^\ddagger\ \left (\ \equiv\
 \Theta^{-1}H^\dagger\,\Theta\ \right )
 \ \ \ \ {\rm in} \ \ \ \ \ {\cal H}^{(S)}\,.
 \label{quasihermi}
 \ee
The auxiliary operator $\Theta=\Theta^\dagger >0$ responsible for
such a change of the definition of Hermitian conjugation in ${\cal
H}^{(S)}$ is called ``metric"~\cite{Geyer}.  {\em All} the other
operators ${\cal O}$ of observables {\em must} be
``cryptohermitian", i.e., Hermitian in the same space,
 \be
 {\cal O} = {\cal O}^\ddagger\ \left (\ \equiv\
 \Theta^{-1}{\cal O}^\dagger\,\Theta\ \right )
 \ \ \ \ {\rm in} \ \ \ \ \ {\cal H}^{(S)}\,.
 \label{quasihermibe}
 \ee
The latter requirement (or a set of requirements if necessary)
specifies the physics of the system in question. In opposite
direction, the set of all the ``hidden hermiticity" requirements
(\ref{quasihermi}) + (\ref{quasihermibe}) may be understood as a
practical recipe for the explicit determination of the correct and
unique metric operator $\Theta$ assigned to given $H$ and ${\cal O}$
\cite{Geyer,SIGMA}.


An immediate consequence of the latter formulation of the theory is
that the two Hilbert spaces ${\cal H}^{(F)}$ and ${\cal H}^{(S)}$
coincide as vector spaces composed of the same wavefunction
elements. The main and only difference between them lies in the
update of the inner product in the latter space. In
Ref.~\cite{SIGMA} a compactified notation using a double-bra symbol
has been recommended,
 \be
 \br \psi_a|\psi_b\kt\ \
 \Longrightarrow\ \
 \br \psi_a|\Theta|\psi_b\kt\ \left (=\br \psi_a|\psi_b\kt^{(S)}
 \ \equiv \ \bbr\psi_a|\psi_b\kt
 \right )\,.
 \label{innerie}
  \ee
Our Hamiltonian $H$ and Schr\"{o}dinger Eq.~(\ref{SEloc}) (or, {\it
mutatis mutandis}, Eq.~(\ref{SEdis}), plus all the other operators
${\cal O}$ of observables) find their physical probabilistic
interpretation in the Hilbert space ${\cal H}^{(S)}$ with a
nontrivial metric $\Theta$. Thus, the internally consistent concept
of the solvability of the model should, in principle, involve {\em
not only} the feasible construction of all the wave functions
$\psi(x)$ and of all the related bound-state spectrum of energies
$E$ {\em but also} the practical feasibility of the assignment of
the metric $\Theta$ to our system. This is a challenging problem,
addressed in our present paper.

\section*{Appendix B: The sample of computation of all the pseudometrics
in the six-dimensional square-well model (\ref{toym}) using the
linear set of Eqs.~(\ref{htotbe}) }


The twelve unknown parameters entering the symmetry-reduced $N=6$
ansatz for the pseudometric
 $$
{\cal Q}^{(6)}_{(a,b,c,d,e,j)}({\lambda})=  \left[ \begin
{array}{cccccc} a&b&c&d&{\it {e}}&{\it {j}}
\\{}b&f&g&h&{\it {k}}&{\it {e}}\\{}c&g&m
&n&h&d\\{}d&h&n&m&g&c\\{}{\it {e}}&{ \it {k}}&h&g&f&b\\{}{\it
{j}}&{\it {e}}&d&c&b&a
\end {array} \right]
 $$
must be shown compatible with the thirty six quasi-Hermiticity
requirements represented by Eq.~(\ref{htotbe}). The reduced set of
six items of these equations (say, number 2, 9, 3, 10 and 4) offers
the affirmative answer. Indeed, this sextuplet of equations
 \ben
 \ba
                        -f + f {\,{\lambda}} + a + a {\,{\lambda}} + c=0\\
                                  -c - c {\,{\lambda}} - m + f + h=0\\
                          -g + g {\,{\lambda}} + b + d=0\\
                        -n - d - d {\,{\lambda}} + g + {k}=0\\
                          -h + h {\,{\lambda}} + c + {e}=0\\
                    -{k} + {k} {\,{\lambda}} + d + {j} + {j} {\,{\lambda}}=0
                    \ea
                    \een
leaves six parameters unconstrained and, with abbreviations
 $$
 U={\frac {c+a
\left( 1+{ \it {\lambda}} \right) }{1-{\it {\lambda}}}}+{\frac
{c+{\it {e}}}{1-{\it {\lambda}}}}-c
 \left( 1+{\it {\lambda}} \right)
 ={\frac {a \left( 1+{ \it {\lambda}} \right) }{1-{\it {\lambda}}}}
 +{\frac {c \left( 1+{\it {\lambda}}^2 \right)}{1-{\it {\lambda}}}}
 +{\frac {{\it {e}}}{1-{\it {\lambda}}}}
 $$
and
 $$
 V={\frac {b+d}{1-{\it {\lambda}}}}+{\frac {d+{\it {j}}\, \left( 1+{\it
{\lambda}} \right) }{1-{\it {\lambda}}}}-d \left( 1+{\it {\lambda}}
 \right)
 ={\frac {j \left( 1+{ \it {\lambda}} \right) }{1-{\it {\lambda}}}}
 +{\frac {d \left( 1+{\it {\lambda}}^2 \right)}{1-{\it {\lambda}}}}
 +{\frac {{\it {b}}}{1-{\it {\lambda}}}}\,
 $$
it specifies the following complete solution of Eq.~(\ref{htotbe})
at $N=6$,
 \ben
   {\cal Q}^{(6)}_{(a,b,c,d,e,j)}({\lambda})=
    \left[ \begin {array}{cccccc} a&b&c&d&{\it {e}}&{\it {j}}
\\{}b&{\frac {c+a \left( 1+{\it {\lambda}} \right) }{1-{\it
{\lambda}}}}&{\frac {b+d}{1-{\it {\lambda}}}}&{\frac {c+{\it
{e}}}{1-{\it {\lambda}}}}&{ \frac {d+{\it {j}}\, \left( 1+{\it
{\lambda}} \right) }{1-{\it {\lambda}}}}&{\it {e}}
\\{}c&{\frac {b+d}{1-{\it {\lambda}}}}&U &
V &{\frac {c+{\it {e}}}{1-{\it {\lambda}}}}&d\\{}d&{ \frac {c+{\it
{e}}}{1-{\it {\lambda}}}}&V&U &{\frac {b +d}{1-{\it
{\lambda}}}}&c\\{}{\it {e}}&{\frac {d+{\it {j}}\,
 \left( 1+{\it {\lambda}} \right) }{1-{\it {\lambda}}}}&{\frac {c+{\it {e}}}{1-{\it
{\lambda}}}}&{\frac {b+d}{1-{\it {\lambda}}}}&{\frac {c+a \left(
1+{\it {\lambda}} \right) }{1-{\it {\lambda}}}}&b\\{}{\it {j}}&{\it
{e}}&d&c&b&a
\end {array} \right]\,.
\label{jafeta}
 \een
At ${j}=0$  this matrix is nine-diagonal. It further
  becomes
 seven-diagonal with ${e}=0$,
pentadiagonal after fixing $d=0$
while the next constraint $c=0$ makes it tridiagonal.
Finally, at $b=0$ we arrive at the simplest, diagonal metric
predicted by Eq.~(\ref{diago}).
%
%
Due to the symmetries of our problem we may infer that, in parallel,
there also exists a very similar antidiagonal metric (\ref{andiago})
containing, again, all units up to the endpoint exceptions.

The similar simplifications of the ``interior" matrix elements in
${\cal P}^{(6)}({\lambda})$ may be noticed to appear in the
tridiagonal and antitridiagonal cases where  the two particularly
elementary special pseudometrics given by the respective
Eqs.~(\ref{bidia}) and (\ref{anbidia}) with $\beta={1-{\it
{\lambda}}}$ are reproduced,
 $$
 {\cal P}^{(6)}_2({\lambda})=
 \left[ \begin {array}{cccccc}
  0&\beta
 &0&0&0&0\\{}
 \beta&0
 &1&0&0&0\\{}0&1&0&1&0&0
 \\{}0&0&1&0&1&0
 \\{}0&0&0&1&0 &\beta
 \\{}0&0&0&0&\beta&0
 \end {array} \right]\,,\ \ \ \
 {\cal P}^{(6)}_5({\lambda})=
 \left[ \begin {array}{cccccc}
                     0&0&0&0&\beta&0\\{}
                     0&0&0&1&0&\beta\\{}
                     0&0&1&0&1&0
 \\{}0&1&0&1&0&0
 \\{}\beta&0&1&0&0&0
 \\{}0&\beta&0&0&0&0
 \end {array} \right]\,.
 $$
The remaining two independent components of the set of pseudometrics
come out as parametrized via $\gamma={({1-{\it {\lambda}}})/({1+{\it
{\lambda}}^2})}$ and $\delta={{1}/({1+{\it {\lambda}}^2})}$ yielding
 $$
 {\cal P}^{(6)}_3({\lambda})=
 \left[ \begin {array}{cccccc}
                     0&0&\gamma&0&0&0\\{}
                     0&\delta&0&\delta&0&0\\{}
                     \gamma&0&1&0&\delta&0
 \\{}0&\delta&0&1&0&\gamma
 \\{}0&0&\delta&0&\delta&0
 \\{}0&0&0&\gamma&0&0
 \end {array} \right]\,,\ \ \ \
 {\cal P}^{(6)}_4({\lambda})=
 \left[ \begin {array}{cccccc}
                     0&0&0&\gamma&0&0\\{}
                     0&0&\delta&0&\delta&0\\{}
                     0&\delta&0&1&0&\gamma
 \\{}\gamma&0&1&0&\delta&0
 \\{}0&\delta&0&\delta&0&0
 \\{}0&0&\gamma&0&0&0
 \end {array} \right]\,,
 $$
in full compatibility with Theorem \ref{hlavni}.

\end{document}